\documentclass[manuscript,screen]{acmart}
\usepackage{subcaption}

\AtBeginDocument{%
  \providecommand\BibTeX{{%
    \normalfont B\kern-0.5em{\scshape i\kern-0.25em b}\kern-0.8em\TeX}}}

\setcopyright{acmcopyright}
\copyrightyear{2018}
\acmYear{2018}
\acmDOI{XXXXXXX.XXXXXXX}

\acmConference[Conference acronym 'XX]{Make sure to enter the correct
  conference title from your rights confirmation emai}{June 03--05,
  2018}{Woodstock, NY}
\acmBooktitle{Woodstock '18: ACM Symposium on Neural Gaze Detection,
 June 03--05, 2018, Woodstock, NY} 
\acmPrice{15.00}
\acmISBN{978-1-4503-XXXX-X/18/06}

\begin{document}


\title[Bringing Utility to NFTs through Multimodal Interaction]{Decentralized, not Dehumanized in the Metaverse: Bringing Utility to NFTs through Multimodal Interaction}

\author{Anqi Wang, Ze Gao, Lik-Hang Lee, Tristan Braud, and Pan Hui}\authornote{All authors, except Lik-Hang Lee, are affiliated with HKUST, China, while Lik-Hang Lee is affilitated with KAIST, South Korea.}




\renewcommand{\shortauthors}{Wang, et al.}

\begin{abstract}

User Interaction for NFTs (Non-fungible Tokens) is gaining increasing attention. Although NFTs have been traditionally single-use and monolithic, recent applications aim to connect multimodal interaction with human behaviour. 
This paper reviews the related technological approaches and business practices in NFT art. We highlight that multimodal interaction is a currently under-studied issue in mainstream NFT art, and conjecture that multimodal interaction is a crucial enabler for decentralization in the NFT community. 
We propose a framework that combines a bottom-up approach with AI multimodal process.
Through this framework, we put forward integrating human behaviour data into generative NFT units, as  "multimodal interactive NFT". Our work displays the possibilities of NFTs in the art world, beyond the traditional 2D and 3D static content.


\end{abstract}




\maketitle

\section{Introduction}

\textit{NFTs (Non-fungible tokens)}\footnote{NFTs -- \url{https://en.wikipedia.org/wiki/Non-fungible_token}} are records of ownership stored on the blockchain. Each token points to a specific piece of content that can be traded by the owner. NFTs have been pitched as a radical centerpiece of Web3.0~\cite{kahan_2021} as a new iteration of the World Wide Web-based on blockchain technology~\cite{edelman_2021}. Compared to virtual currencies circulating in the Web3.0 world such as Bitcoin or Ethereum, NFTs can represent any type of content, including physical land property, digital asssets, music, and art. Most of the burgeoning NFTs are known for their digital art collections~\cite{valeonti2021crypto}. Recently, the development of data multimodality has introduced methods with significant progress in text-to-image generation~\cite{ramesh2021zero,radford2021learning}. For example, OpenAI has launched DALL·E after CLIP, an AI system for generating images from textual descriptions using a dataset of text-to-image pairs, incorporating multimodal processing~\cite{openai_2021,ramesh2021zero}. 
This text-to-image generation reveals the possibility of combining data multimodal techniques with the artistic process. 
A large body of work builds on existing text-to-image data multimodality approaches, leading to multiple open-source projects~\cite{esser2021taming,crowson2022VQ}. 
There is also a proliferation of business cases for such works, e.g., supporting people to generate artworks using text input for AI processing~\cite{reddit2022bigsleep}. Existing mainstream methods and models rely on \textit{GANs}\footnote{GANs -- \url{https://en.wikipedia.org/wiki/Generative_adversarial_network}} using textual latents~\cite{openai_2021}. 
With text and images being considered as the primary NFT forms, there is potential for text-to-image generation and data multimodality to integrate the NFT world~\cite{regner2019nfts}.


However, NFTs under their popularly adopted forms are considered too monolithic in form and lack interactivity~\cite{pacella_2021,hashkey2021combing,ello2022blockchain,russo_2021}. Some studies suggest that future NFTs should be more interactive and responsive to human perception in real-time~\cite{chainlink_2021}. 
A multimodal approach has the potential to address this dilemma, paving the way for novel interaction techniques to integrate with NFTs in the future.
As a matter of fact, many existing NFT commodities already propose some degree of interaction and multimodal processing, with commerce at its forefront~\cite{ordano2017decentraland,chainlink_2021}. 
However, most projects target a single use-case, and there is a lack of comprehensive frameworks describing multimodal interaction with NFTs.

This paper aims to contribute a comprehensive framework of multimodal interaction with NFTs, with AI and generative techniques at its core. We first investigate the state-of-the-art intersection of interactive NFTs and multimodality and propose that multimodal real-time interaction is one of the future directions for NFTs.
We then analyze multimodal interaction in NFTs through two case studies. One is an intelligent NFT with multimodal AI (artificial intelligence) processing; another is an image-based NFT created entirely by the users according to text information. We note that these two types of NFTs can complement each other through data exchange.
We then propose a comprehensive framework for creating multimodal interactive NFTs that can respond in real-time, combining a bottom-up approach with text-input-to-image multimodal AI. This new approach and pipeline can serve as a novel paradigm for 
interactive NFTs. Finally, we clarify the significance of multimodal processing in art and culture. 
This proposed framework sheds light on potential NFTs' use of multimodal data for interactive, generative, and evolutive content creation and ownership.


\section{Related Works}

\subsection{NFTs Utility and Interactivity}


Gary Ma, Chief Operating Officer at digital collectibles developer Epik stated that the NFT market would lose a lot of potential users unless users found utility in NFT assets~\cite{UknowAuth2022nftfuture}. Users would thus be able to engage with NFT assets through form of games and applications.
Psychologist Mark McKinley further describes user engagement with NFTs through four factors: Investment, Social Expansion, Sense of Self, and Regain Control~\cite{shah_2021}.

In terms of Investment, NFTs have seen many large swings in value from an investment and value perspective, for instance, some NFTs rising 10-100 times in price and plummeting 92\% in a few weeks \cite{dean_2022}. This lack of utility has led to an ``economic collapse," resulting in a relatively limited circulation of NFTs. The CACAU NFT Valuation Model proposed that NFT Value comprises two core components: Narrative Value and Utility Value \cite{teng_2021}. Most NFTs focus only on the narrative value, while the utility deficit may be one of the causes of the problem. 

Social Expansion, Sense of Self, and Regain control all correlate to social attributes and how stakeholders participate in NFT projects. 
The existing static, non-interactive NFTs do not bring much more experience to the user. Most NFTs lie in the owner's wallet address or only as a ticket to the NFTs' community~\cite{regner2019nfts}. 
They do not bring utility value to the user: NFTs' production processes or interactions do not link to users' behavior and actual social activities. The lacking of utility value leads to a reduced sense of self and regained control. Even if the NFTs give the colletors unique ownership, the sense of incentive gained by the purchase can quickly fade. From the perspective of long-term holding, collectos will lose their sense of belonging and uniqueness. NFTs are expecting more radically new forms that can build a deeper relationship with users, such as the gaming industry and virtual events~\cite{wang2021non}.

Together, these four factors, Investment, Social Expansion, Sense of Self, and Regain Control, help us understand why NFTs have not reached a broad audience yet. Many studies have proposed to introduce interaction to the current static, non-interactive NFTs to increase their utility ~\cite{russo_2021,mirrorworld,hashkey2021combing,ello2022blockchain,chainlink_2021,wang2021non}, eventually making the value of NFT more sustainable.

\subsection{Multimodality in Arts}

\begin{figure}
\centering

   
\begin{subfigure}[t]{.49\textwidth}
\centering
  \includegraphics[width=0.75\textwidth]{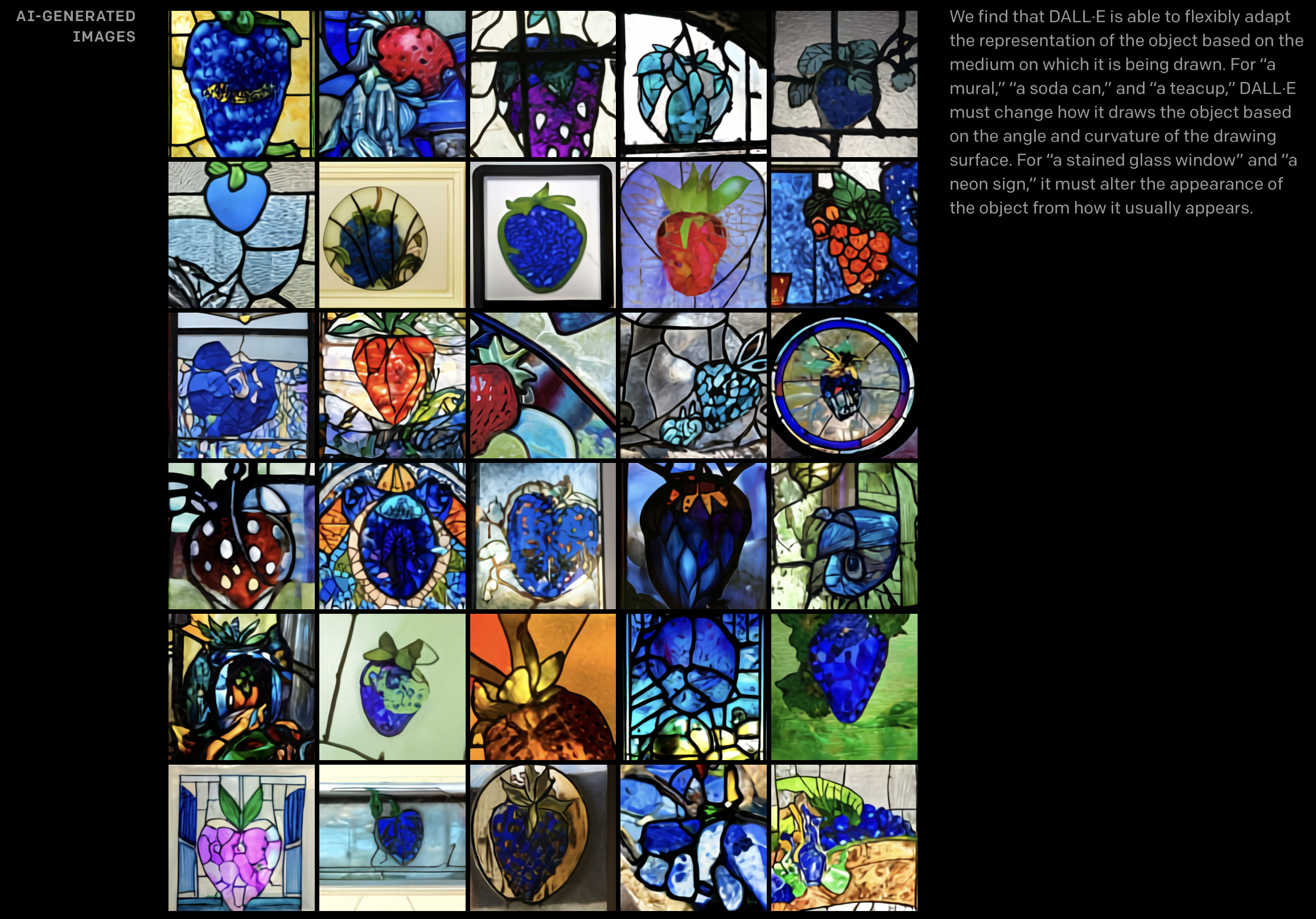}
  \caption{Text Prompt: “A stained glass window with an image of a blue”, AI generates images of different styles.}
  \end{subfigure}
  \hfill
  \begin{subfigure}[b]{.49\textwidth}
  \centering
  \includegraphics[width=0.75\textwidth]{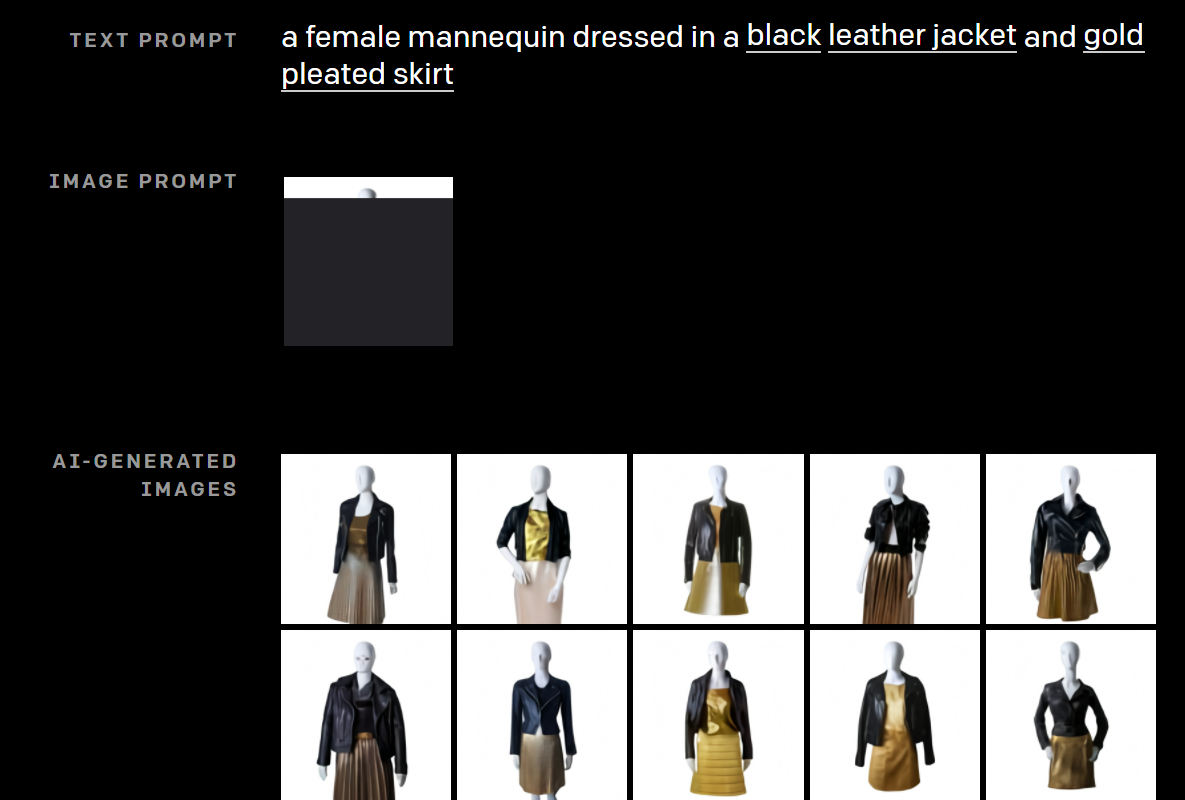}
\caption{Generate different styles of female models wearing fashion.}
  \end{subfigure}
\caption{DALL·E's application, source: OpenAI blog, \url{https://openai.com/blog/dall-e/}.}
  \label{fig:1}
  
 \end{figure}

Many multimodal used for generative art creation~\cite{twi2022multi} and artistic analyses~\cite{liu2022design} and text-and-image approaches combining AI process have appeared in many artistic cases. \textit{DALL-E}\footnote{DALL-E -- \url{https://openai.com/blog/dall-e/}}, one of those tools, aims to generate images based on text descriptions (See Fig.~\ref{fig:1}). In contrast, \textit{CLIP}\footnote{VQGAN+CLIP -- \url{https://colab.research.google.com/github/justinjohn0306/VQGAN-CLIP/blob/main/VQGAN2BCLIP(Updated).ipynb}} (contrastive language image pre-training) aims to associate text and images more robustly than current AI models. OpenAI builds both. The essence of this type of multimodal framework for text-to-image synthesis is pre-train the textual information on a large-scale image on the Internet. It corresponds to the image's content expressed in response and then outputs an image result that fuses with the textual information and the original image. 
The user can change some text input parameters to see how the output is affected, such as the number of iterations and the adjectives of various attributes. 

The usages of these sophisticated models provide us with unprecedented analysis and creativity, especially in artistic creation and interaction, and many studies have demonstrated their potential \cite{twitter2022rivershavewings}. The \textit{Portrait of Edmund Bellamy} created by AI already sells at high prices~\cite{wang2019value,yu2020research}, and generated artwork by multimodal may have more commercial value because the input has multiple datasets having better interaction with users.
It also indirectly reveals its artistic value. Therefore, integrating DALL·E, an artificial intelligence generation method with a text-to-image process, as an input method can enable these interactive co-creation methods of NFTs.

\subsection{Multimodal Interaction with NFTs}
 Interactive and intelligent NFTs represent one typical promising NFTs that provide interaction methods to collectors and have their iteration with the AI process. The following section introduces how multimodal data can combine interactive and intelligent NFTs from a multimodal perspective.

Some pioneering NFT cases are using multimodal techniques, mainly focusing on multimodal AI or multimodal data input techniques for AI. These types of NFTs can be called ``Interactive, reactive, and intelligent NFTs" in terms of users' inputs, environments, or other sources and altering the condition or aesthetics of an original piece of content~\cite{pacella_2021}. Therefore, multimodal techniques can serve as an input method for this type of NFT to achieve the so-called dynamic NFTs.  We consider two case studies that leverage generative AI to create interactive and intelligent NFTs.

\textit{Humans.ai}\footnote{Humans.ai -- \url{https://humans.ai/}}, using AI technology to create voice NFTs that encapsulate a person's voice in each NFT, creates a unique voice using the interaction of input text~\cite{humans.ai_2022}. It focuses on a multimodal approach. Human.ai treats the NFT as a dynamic asset from one mode to another where the AI processes text-to-speech interactions. Multimodality in Human.ai accomplishes the fusion of textual data with the speech of unique individuals.

\textit{Mirror World NFT}\footnote{Mirror World NFT -- \url{https://mirrorworld.fun/}},  created in September 2021, is the first NFT project to apply more AI models to Web 3 communities. The Mirror is a collection of 11,000 unique AI Virtual Beings. Each Mirror can be upgraded and co-create narratives by talking with the collector. Each Mirror can be upgraded, also offering a series of rights in the future games~\cite{mirrorworld}. This process demonstrates Mirror's unique concept of AI generation and support for conversational interaction with the collectors. Multimodality translates text into multimodal processing, then presents in a real-time responsive manner to the visual characters displayed in NFT.

This multimodal AI is packaged under different names in different projects. For instance, Mirror World's multimodal AI is called "Soul", an artificial intelligence dialogue chip. Their essence is the engagement of arts based on a multimodal approach.
Multimodal AI in NFTs means that the owner cultivates their own NFTs, embedding a personalized narrative and unique ownership into the generated NFTs. This multimodal input brings multiple interactions, giving NFTs greater creativity and higher human value.


\section{Description}

\begin{figure}
    \centering
    \includegraphics[width=0.7\textwidth]{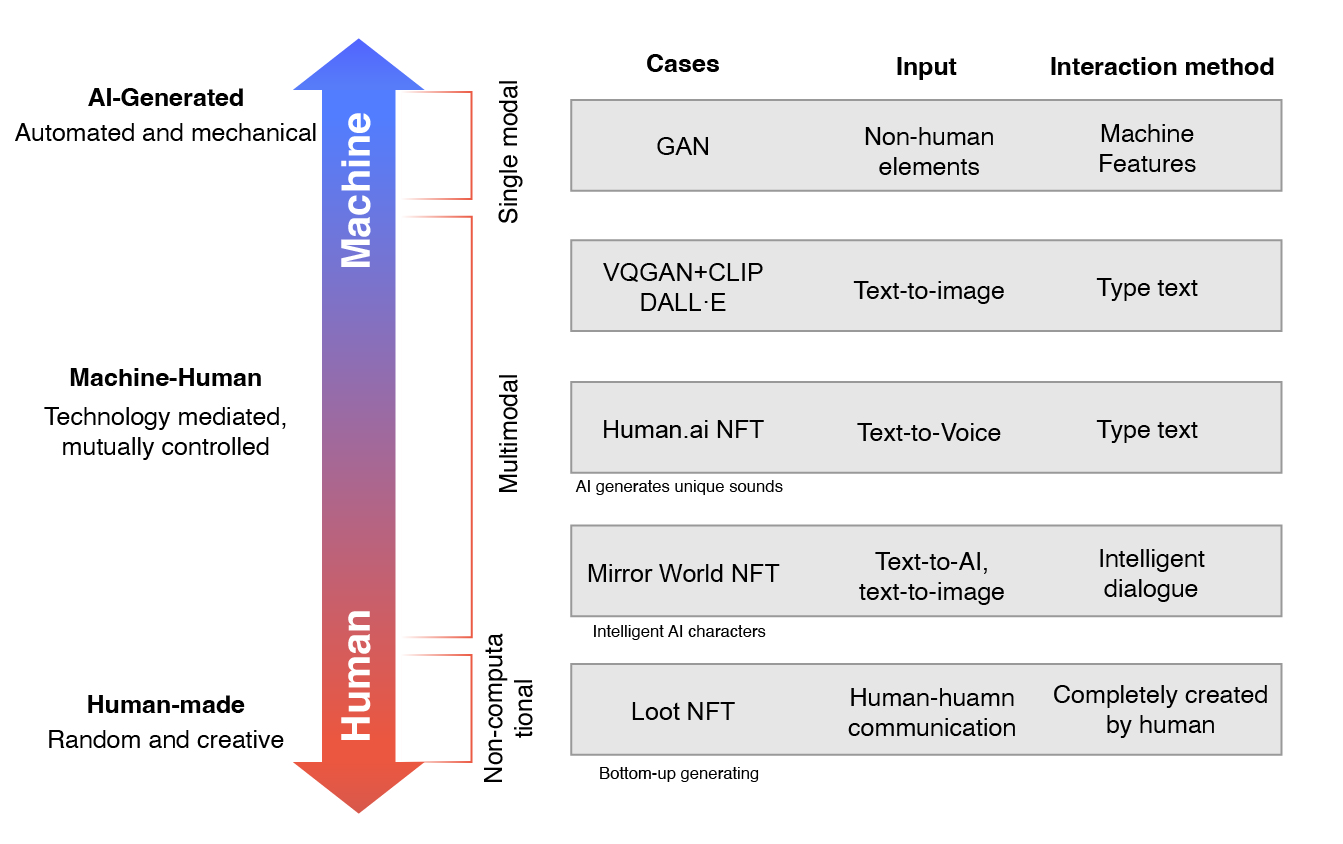}
    \caption{Diagram of a human-machine continuum in multimodal NFTs. }
    \label{fig:continuum}
\end{figure}

In the previous sections, we shed light on the problem of insufficient multimodality in NFTs that leads to a problem in user participation. Meanwhile, there is a lack of unified pipelines or frameworks to introduce such multimodality to the productors.
This section proposes a continuum theory for multimodal NFTs regarding the relationship between Humans and Machines. The degree of human-machine interaction represents
the mode and extent to which multimodal data integrates with human factors (See Fig~\ref{fig:continuum}). Consequently, we gain insight into how we should construct future NFTs. We then propose a novel interaction method for NFTs based on this theory, which links multimodal data with generative AI into the existing text-to-image interaction method for NFTs (See Fig~\ref{fig:framework}).

\subsection{Bottom-up Approach}
\begin{figure}
    \centering
    \includegraphics[width=0.75\textwidth]{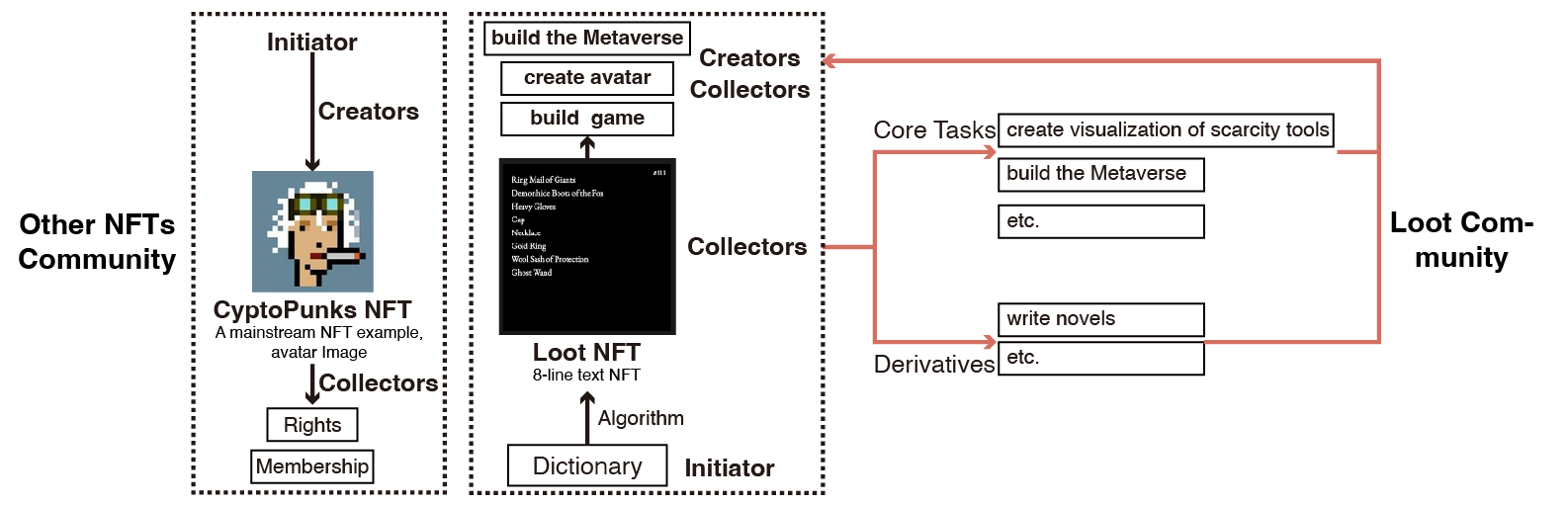}
    \caption{Loot Metaverse, a bottom-up approach in the community compared to other NFTs community}
    \label{fig:loot}
\end{figure}

Before discussing the human-machine continuum that leads to emerging protocol paradigms for NFTs, it is crucial to understand the bottom-up approach. The bottom-up approach is where the user is entirely free to create NFT images based on specific information (e.g, Loot's pre-set 8 lines of text). This approach is a radical milestone that stands on the ``human" side, without any involvement of algorithms or AI, and is entirely determined and created by users.

The bottom-up approach is a method of creating an NFT that has the potential to revitalize the NFT-user relationship directly~\cite{wang2021non}. \textit{Loot NFT}\footnote{Loot NFT -- \url{https://lootnft.io/}} project is one of such communities that is open-sourced and self-organized from the ground up (See Fig.~\ref{fig:loot}). A single Loot NFT is a list of items consisting of only eight lines of text. Users create NFT images based on these eight lines of text, even though the Loot project does not provide any stylized images or official roadmap. However, the project organized a community with clear goals and system architecture within the first five days of its release. Such a bottom-up approach essentially establishes a more profound emotional logic with the users and enhances the relationship between the users and the NFT community.
Even if the project's initiators handed out the roadmap and goals of Loot NFT to the community, most participants still lack specific strong incentives as other interactive elements are missing in the community. There are multiple limitations to the bottom-up approach, such as the sparse influence outside the community, the loss of integrated governance, and the efficiency in goal achievement.




\subsection{Human-Machine Continuum with Multi-modal and Bottom-up Approach}

We propose a relational diagram of the human-machine continuum for multimodal interaction NFTs. We form a theory of balancing humans and machines divided into three categories: human-made, multimodal, and single-modal approaches, centered around the degree of human factor involved (see Fig.~\ref{fig:continuum}). We believe that combining the bottom-up approach with multimodal data can address the limitations of each other. Thus, it is significant to analogize these approaches in these different cases and establish continuity to explore the degree and method of involvement. For the relationship between human factors and the multimodal process, the approaches from the human side represent random and creative forces made entirely by humans. The approaches from the machine-side mean the use of multimodal data as a method of generating art after the mediation by the machine. 
The bottom-up approach and multimodal data can solve each other's problems. For example, multimodal data limits the disorderly nature of bottom-up human creativity with set input parameters and quantity, transformation process, and output. This process effectively standardizes content production, increasing efficiency and clarifying its purpose.
Thus, to clarify this approach, Fig.~\ref{fig:continuum} shows the different multimodal NFT cases and multimodal tools in the art mentioned in the previous case studies according to three different classifications. Moreover, it summarizes the degree of human-machine involvement in the different cases according to the approach to their support of user input, process, and interaction methods.

\subsection{Multimodal NFTs' Proposal}

\begin{figure}
    \centering
    \includegraphics[width=0.95\textwidth]{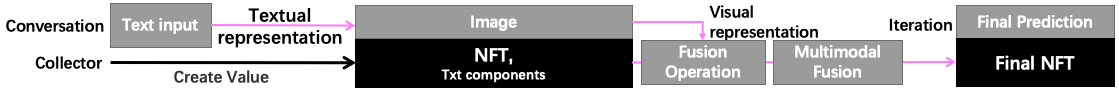}
    \caption{Framework for generating interactive NFT with AI multimodality}
    \label{fig:framework}
\end{figure}


We choose a multimodal text-to-image input approach in which the user can type text to interactively iterate over existing NFTs.
 Multimodality is intrinsic to human communication and texts and consists of an assortment of signs that operate in communicative contexts~\cite{lahdesmaki2022multimodality}. Language is characterized in psychology as the most representative of a person's personality~\cite{mairesse2007using}. At the same time, vision is undoubtedly the most focused sensory experience in NFTs~\cite{nadini2021mapping}. Therefore, we have chosen both linguistic interaction with text and visual interaction as the most direct and representative ways of interaction.

This framework spreads over two layers. The underlying layer consists in a dictionary provided as a text file by the project. This dictionary fits the worldview and concepts of the NFT project. The words in the dictionary can form various phrases according to algorithms defined by the project. These form the minor units that make up the NFTs, such as the teeth, necklace, and pupil color of an avatar NFT. In this paper, we take the user-generative approach introduced by Loot NFT that supports the bottom-top construction of NFTs, where users can create and mint their own NFT image (See Fig.~\ref{fig:framework}"Image").
This first layer sets the foundations for the multimodal approach.
The second layer is a text-to-image multimodal data processing system that incorporates the functionality of NFT and user-AI dialogues. The user enters the text of the conversation on the NFT interface, and the ``Image" of the NFT built on with first layer changes dynamically. The information contained into the ``Image" and the text information of the ``Text Input" typed by the user together form a dual data input, which fuses with the multimodal processing of AI. It targets the given task through dialogue-based and NFT interactions. This technique can map objects that occurred in different modalities or texts expressed in different languages into a joint semantic space. 
This fusion of dual dataset input and AI multimodal processing supports the input qualification of custom creations with a minimum unit of bottom-up. The processing and fusion between visual image and conversational text limit the variable approach to interactive NFT in a way that maximizes tolerance and fit. It can support highly personalized narratives in which each NFT has the most intense individual characteristics of its collectors.


\section{ACHIEVING IMPACT}
Text-to-image interaction presents a novel and emerging form of human-computer interaction for media creation~\cite{liu2022design}. It operates as a blend of human factors and machine processing, balancing human factors with machine data processing. Multimodality as an underlying technology interacts with art and culture to build better representations in Web3.0 and leads to a more humanistic society. The following are the implications of social interaction, social expansion, and self-actualization.

\textbf{Social Interaction}
How multimodality can build a more humane society is a highly debated topic~\cite{oviatt2021technology}. In the broad view, multimodality involves leveraging human communication styles through verbal and nonverbal communication. Understanding human-machine interaction is fundamental to the long-term pursuit of powerful and natural multimodal interfaces. Communication using text as input is only the first step. Integrated human behavior data, such as nonverbal communication (step count, emotions, facial expressions, body posture, gesture, and eye gaze~\cite{10.1145/1027933.1027942}), to generative visualization of NFT units, not only affect the future of NFT but are also critical to human social interaction.

\textbf{Social Expansion}
After generating interactive NFTs through AI-driven multimodal interaction, the intelligent conversational format and fusion of images will bring the near-infinite nature of NFT creation and the strategy of user decisions and aesthetic choices. It will create a free and open user-centric Web3.0 environment, opening the door to experiments in digital improvisation. 
Its bottom-up construction logic will continue to support our understanding of what collaborations are possible in this process and future Web3.0 worlds.

\textbf{Sense of Self}
The self-training process translates the users' characteristics into the NFTs. Through such self-actualization~\cite{maslow1971self},  the visual elements of NFTs can display the personality and even preferences of the owner. For instance, Scatter Lab launched a Lee-Luda in 2020, which is a defunct Korean female AI chatbot model~\cite{lim2022review}. ``She" gains rising popularity because "She" has a tight emotional connection to each user. By conversing with a user, the AI model integrates elements of the user's personality, progressively creating a personalised experience. Emotional growth and authorship give users more unique experiences and feelings to achieve self-actualization.

\section{Conclusion}
NFTs can build a better Web3.0 marketplace and network, bring user experience, and start a highly active and cultural community~\cite{regner2019nfts,Lee2021WhenCM}. They have the power to enhance these benefits in an expanded, more interactive form~\cite{pacella_2021,regner2019nfts}. So far, OpenSea is the largest trading marketplace~\cite{hoogendoorn_2021}, and users earn NFTs primarily by buying and selling on the marketplace. However, we will soon see an evolution that includes earning them through user behavior or other participation incentives~\cite{pacella_2021}. For example, similar to the existing way of self-minting NFTs, users' self-participation enhances their emotional interaction with NFTs rather than directly buying them, creating the value of their commodity. Currently, existing NFT interaction and utility are insufficient, and multimodal data can allow people to interact with NFT for the sake of enhanced utility. 
This paper investigates the impact of interactive NFTs and multimodality in arts and proposes a novel framework -- multimodal production methods for interacting NFTs with human input and creation. 
We call for more research efforts on interactive and intelligent NFTs based on the proposed bottom-up generation approach, the integration of dialogue-based text, and AI multimodality, which disrupt publishers' production and selling of NFTs and provide a new landscape on NFTs-human interaction. 

\bibliographystyle{abbrv}
\bibliography{sample-base} 

\end{document}